\documentclass[letterpaper,10pt,twocolumn,superscriptaddress,floatfix]{revtex4-1}
\usepackage[utf8]{inputenc}
\usepackage{amsfonts,amssymb,amsmath,amsthm,graphicx}
\usepackage{url}
\usepackage{dsfont}
\usepackage{bbm}
\usepackage[usenames,dvipsnames]{xcolor}
\usepackage{nicefrac}
\usepackage{natbib}
\usepackage{setspace}
\usepackage{subfigure}
\usepackage{comment}
\usepackage{enumitem}
\usepackage{hyperref}

\begin{document}

\title{Sub-ns timing accuracy for satellite quantum communications}

\author{Costantino Agnesi}
\affiliation{Dipartimento di Ingegneria dell'Informazione, Universit\`a degli Studi di Padova, via Gradenigo 6B, 35131 Padova, Italy}
\affiliation{Istituto Nazionale di Fisica Nucleare (INFN) - sezione di Padova, Italy}

\author{Luca Calderaro}
\affiliation{Dipartimento di Ingegneria dell'Informazione, Universit\`a degli Studi di Padova, via Gradenigo 6B, 35131 Padova, Italy}
\affiliation{Istituto Nazionale di Fisica Nucleare (INFN) - sezione di Padova, Italy}
\affiliation{Centro di Ateneo di Studi e Attivit\`a Spaziali ``G. Colombo", Universit\`a degli Studi di Padova, via~Venezia~15, 35131~Padova, Italy}

\author{Daniele Dequal}
\affiliation{Matera Laser Ranging Observatory, Agenzia Spaziale Italiana, Matera, Italy}

\author{Francesco Vedovato}
\affiliation{Dipartimento di Ingegneria dell'Informazione, Universit\`a degli Studi di Padova, via Gradenigo 6B, 35131 Padova, Italy}
\affiliation{Istituto Nazionale di Fisica Nucleare (INFN) - sezione di Padova, Italy}
\affiliation{Centro di Ateneo di Studi e Attivit\`a Spaziali ``G. Colombo", Universit\`a degli Studi di Padova, via~Venezia~15, 35131~Padova, Italy}

\author{Matteo Schiavon}
\affiliation{Dipartimento di Ingegneria dell'Informazione, Universit\`a degli Studi di Padova, via Gradenigo 6B, 35131 Padova, Italy}
\affiliation{Istituto Nazionale di Fisica Nucleare (INFN) - sezione di Padova, Italy}

\author{Alberto Santamato}
\affiliation{Dipartimento di Ingegneria dell'Informazione, Universit\`a degli Studi di Padova, via Gradenigo 6B, 35131 Padova, Italy}

\author{Vincenza Luceri}
\affiliation{e-GEOS SpA, Matera, Italy}

\author{Giuseppe Bianco}
\affiliation{Matera Laser Ranging Observatory, Agenzia Spaziale Italiana, Matera, Italy}

\author{Giuseppe Vallone}
\affiliation{Dipartimento di Ingegneria dell'Informazione, Universit\`a degli Studi di Padova, via Gradenigo 6B, 35131 Padova, Italy}
\affiliation{Istituto Nazionale di Fisica Nucleare (INFN) - sezione di Padova, Italy}

\author{Paolo Villoresi}
\email{paolo.villoresi@dei.unipd.it}
\affiliation{Dipartimento di Ingegneria dell'Informazione, Universit\`a degli Studi di Padova, via Gradenigo 6B, 35131 Padova, Italy}
\affiliation{Istituto Nazionale di Fisica Nucleare (INFN) - sezione di Padova, Italy}

\begin{abstract}
Satellite quantum communications  have rapidly evolved in the past few years, culminating in the proposal, development and deployment of satellite missions dedicated to Quantum Key Distribution and the realization of fundamental tests of quantum mechanics in space. However, in comparison with the more mature technology based on fiber optics, several challenges are still open, such as the capability of detecting with high temporal accuracy single photons coming from orbiting terminals. Satellite Laser Ranging, commonly used to estimate satellite distance, could also be exploited to overcome this challenge.  For example, high repetition rates and low background noise can be obtained by determining the time-of-flight of faint laser pulses that are retro-reflected by geodynamics satellites and then detected on Earth at the single-photon level. Here we report on a experiment achieving a temporal accuracy of about 230~ps in the detection of an optical signal of few photons per pulse reflected by satellites in medium Earth orbit, at a distance exceeding 7500 km by using commercially available detectors. Lastly, the performance of the Matera Laser Ranging Observatory are evaluated in terms of detection rate and signal to noise ratio for satellite quantum communications.
\end{abstract}

\maketitle

\section{Introduction}
Satellite quantum communications (SQCs) {  have} seen rapid progress since they were first proposed in 2002~\cite{Rarity2002}. Feasibility studies~\cite{Villoresi2008, Bonato2009, Tomaello2011} have led to proof-of-principle experiments~\cite{Wang2013, Yin2013, Vallone2015, Vallone2016, Dequal2016, Pugh2017, Gunthner2017, Calderaro2018} which paved the way for the deployment of a fully-functioning Quantum Key Distribution (QKD) satellite~\cite{Pan2014, Liao2017, Yin2017_PRL, Liao2018} and the realization of fundamental tests of quantum mechanics in space~\cite{Yin2017_Science, Ren2017, Vedovato2017} (see the reviews~\cite{Bedington2017, Khan2018, Agnesi2018, Pan2018} for further details). SQC is the enabling technology for a global scale quantum internet~\cite{Kimble2008}, since fiber-optic links are limited to a few hundred kilometres due to an exponential signal attenuation with distance~\cite{Yin2016,Boaron2018_record}, and quantum repeaters are still far from use in practical implementations~\cite{Sangouard2011}.

The goal of QKD is to distill a secret key between two distant parties with security that can be proved to be unconditional. Such level of security cannot be reached with classical communication schemes that rely on computational hardness assumptions, many of which will be undermined as quantum computers become more mature \cite{Vandersypen2001, Politi2009}.  As of today, QKD is the most advanced application of quantum communications, having both a solid theoretical framework and widespread experimental implementation, even for commercial purposes \cite{Gisin2002, Scarani2008}.   
However, despite the recent results of SQC, several technical challenges  still remain, mainly in terms of achievable repetition rate and secret key rate, as well as link stability. In fact, fiber-optics based implementations are still the benchmark for repetition rate~\cite{Boaron2018, Grunenfelder2018}, obtained secure key rate~\cite{Lucamarini2013, Islam2017} and stability~\cite{Dynes2012, Yoshino2013}. On the other hand, recent satellite-based QKD experiments are limited to 100~MHz repetition rate, few kilohertz of secure key rate with Low Earth Orbit (LEO) satellites,  2~ns detection window, and link duration of few minutes~\cite{Liao2017}.

Satellite Laser Ranging (SLR), a technique of fundamental importance for space geodesy~\cite{Pearlman2002, Luceri2015}, can play an instrumental role in the development of SQC. By determining the satellite orbit with millimeter accuracy~\cite{Degnan1993}, many orbital parameters can be estimated and used to predict and observe phenomena caused by the motion of the satellite (see~\cite{Vallone2016, Vedovato2017}).  Furthermore, by determining the time-of-flight of light signals  with high precision, stringent temporal filtering can be employed, obtaining a lower background noise and
enabling the realization of satellite quantum communications with high repetition rates.

\begin{figure}[!htb]
\centering\includegraphics[width=0.7\columnwidth]{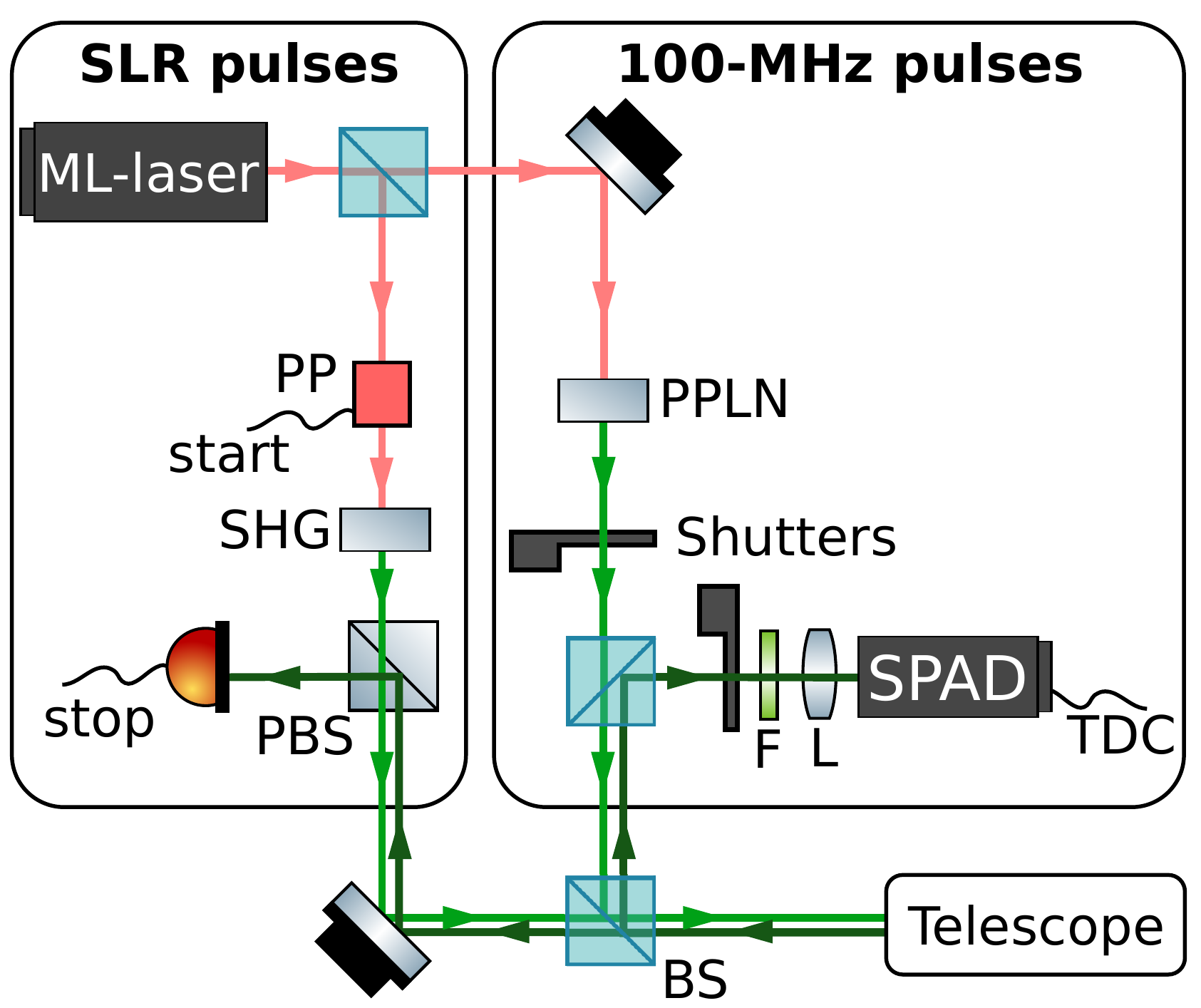}
\caption{Scheme of the experimental setup as described in Section~\ref{sec:Setup}. }
\label{fig:Setup}
\end{figure}

In this article, we report on the advances in the detection and timing techniques implemented at the Matera Laser Ranging Observatory (MLRO) of the Italian Space Agency~\cite{Bianco2018}, that allow a temporal accuracy of few hundreds of picoseconds in the detection of an optical  pulse with few photons reflected by SLR satellites in medium Earth orbit (MEO) at a distance exceeding 7500 km using commercially available single-photon detectors and time tagging electronics. Lastly, we estimated the performances, in terms of detection rate and signal to noise ratio (SNR), of SQC with MEO satellites with our setup.

\section{Experimental Setup}
\label{sec:Setup}
The experiment was performed at the MLRO ground station by using the setup sketched in Fig.~\ref{fig:Setup}. This setup was also used in our recent study  on the feasibility of quantum communications from Global Navigation Satellite System~\cite{Calderaro2018}.

\subsection{Generation and detection of the SLR pulses}
The MLRO observatory is a SLR station equipped with a mode-locking Nd:YVO$_4$ laser oscillator (ML-laser), operating at 1064~nm  with 100~MHz repetition rate and paced by an atomic clock. The SLR pulses, with wavelength $\lambda = 532$ nm, $\approx100$~mJ of energy and 10~Hz repetition rate, are obtained by selecting one seed pulse every $10^7$ with a pulse-picker (PP), which is then amplified and up-converted via a second-harmonic-generation (SHG) stage. The SLR pulses are sent to the targeted satellites, which are equipped with corner-cube retroreflectors (CCRs)~\cite{Degnan1993}, using the 1.5-m diffraction-limited Cassegrain telescope of MLRO.

For this study, the Laser Geodynamic (LAGEOS-I and LAGEOS-II) and Beacon-C satellites were chosen. The LAGEOS satellites, jointly lunched by NASA and ASI in 1976 and 1992 respectively, have been used to study perturbations in Earth's gravitational field~\cite{Yoder1983}, to observe the Lense-Thirring Effect~\cite{Ciufolini1998},  as well as in the first demonstration of SQC with MEO satellites~\cite{Dequal2016}. The Beacon-C LEO satellite, launched by NASA in 1965 for ionospheric research and geodesy, has also been used to study time-bin encoding for SQC~\cite{Vallone2016} and  to extend Wheeler's delayed-choice experiment to space~\cite{Vedovato2017}.

After the reflection by the orbiting terminals, the SLR pulses are collected by a fast analog micro-channel plate detector (Hamamatsu R5916U-50) placed after a polarizing beam-splitter (PBS) used to separate the transmitted beam from the received one. A dedicated time-tagger with picosecond accuracy recorded the start and stop signals generated by the PP and the detector respectively. The single-shot measurement of the satellite distance is then estimated from the time-difference of these two signals, i.e. the  \textit{round-trip-time},  with an average normal point accuracy of $3$~mm (root-mean-square) for LAGEOS~\cite{MLROprecision}.

\subsection{Generation and detection of the 100-MHz pulses}
A setup dedicated to study the feasibility of SQC  is implemented in parallel to the SLR system. The same ML-laser is used to produce a 100-MHz pulse-train with wavelength $\lambda = 532$ nm, $\approx$1 nJ of energy and 55~ps of pulse duration at full-width-half-maximum (FWHM) by exploiting  a SHG stage given by a 50 mm long periodically poled lithium niobate non linear crystal from HC Photonics. This beam, synchronized with the SLR pulse-train, is combined with the outgoing SLR pulses by using a 50:50 beam splitter (BS) and the two light beams are sent to the targeted satellites.

The receiving apparatus of the 100-MHz beam is comprised of a 50:50 BS to separate the outgoing and incoming beams, a $3$~nm FWHM spectral filter (F) with transmission band centered at $532$~nm, a focusing lens (L) and a silicon single photon avalanche detector (SPAD), provided by Micro-Photon-Devices Srl, with  $\approx50\%$ quantum efficiency, $\approx400$~Hz dark count rate and $40$~ps of jitter. The time  of arrival of the returning photons is recorded with $1$~ps resolution by the quTAG time-to-digital converter (TDC) from qutools GmbH. This represents a substantial upgrade in terms of timing jitter and detection efficiency with respect to the detection system used at MLRO for other SQC experiments~\cite{Vallone2015,Vallone2016,Vedovato2017}. In fact, the previous detection system was composed of single-photon photomultipliers (Hamamatsu H7360-02) with an active area of 22~mm diameter, $\approx10\%$ detection efficiency, $\approx50$~Hz dark count rate and 500~ps of timing jitter, and a quTAU TDC, also from qutools GmbH, with 81~ps of temporal resolution. Such detection system allowed a temporal accuracy slightly larger than 1ns 
in the detection of an optical pulse with few photons reflected by SLR satellites
(the FWHM reported in~\cite{Vallone2015, Vallone2016, Dequal2016, Vedovato2017}  is $\simeq 1.2$~ns).

\begin{figure*}[t]
\centering
\includegraphics[width=\textwidth]{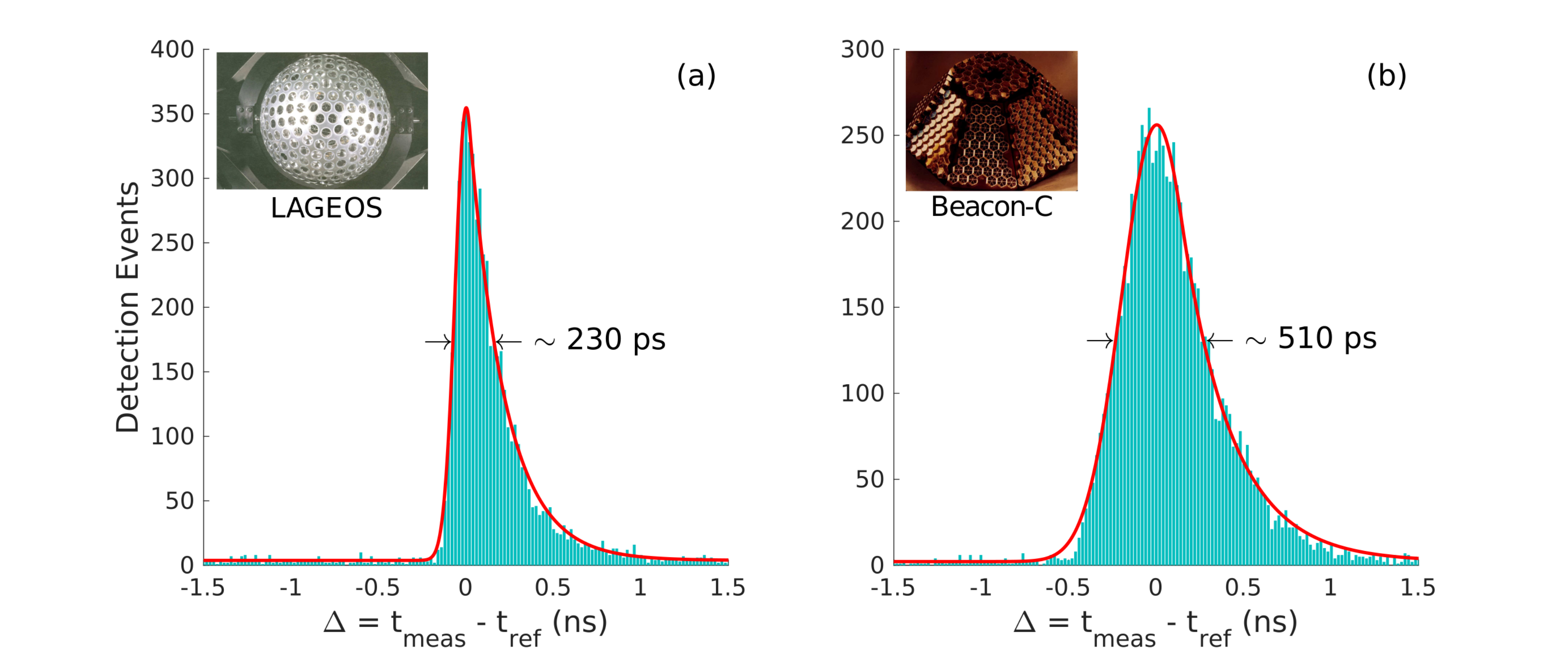}
\caption{(a) Detection histogram for 5 seconds of LAGEOS-I passage. A FWHM of 230~ps is observed. Similar results were also obtained with LAGEOS-II. (b) Detection histogram for 5 seconds of Beacon-C passage. A FWHM of 510~ps is observed. The broadening is due to the shape of the satellite. Both histograms are obtained with a binning of 20~ps. The red curves are obtained by fitting the data with Eq.~\eqref{eq:detResp}. The inset photographs of the satellites are courtesy of the \href{https://ilrs.cddis.eosdis.nasa.gov/}{International Laser Ranging Service (ILRS)}.   }
\twocolumngrid
\label{fig:detectionHist}
\end{figure*}

\subsection{Transmission and reception procedure}
We implemented a procedure to separate the transmitting and receiving phases by using two mechanical shutters, dividing the experiment in slots of equal length. 
In the first half of the slot, the transmitting (receiving) shutter is open (closed) and the 100-MHz pulses are transmitted. Vice-versa, in the second half of the slot the receiving (transmitting) shutter is open (closed) and the 100-MHz pulses coming from the satellite can be detected. 
Since the round trip time of photons reflected by LAGEOS satellites ranges from $40$~ms to $55$~ms, each slot has a duration of $100$~ms.
In particular, each slot starts with the SLR start signal at $t=0$~ms. The 100-MHz pulses are sent to the satellite from $t=0$~ms to $t=40$~ms, opening the shutter placed in the transmission path. At $t=50$~ms the receiving shutter opens the receiving path until $t=90$~ms.

\section{Results}
Passages of LAGEOS-1, LAGEOS-2 and Beacon-C on October 16, 2017 were chosen for this study. The LAGEOS satellites were adopted since their optical transfer functions have been extensively characterized~\cite{Arnold1978, Minott1993}. Furthermore, the cannon-ball design guarantees that the temporal profile of the retro-reflected pulse suffers less temporal spread than by a flat panel array of CCRs~\cite{Degnan1993}. These characteristics render the LAGEOS satellites ideal to characterize the timing performance of the  new SQC detection system implemented at MLRO (see Section~\ref{sec:detTiming}), and to characterize the quantum channel (see Sections \ref{sec:detRate} and \ref{sec:muSat}).

\subsection{Resolving photons arrival time with sub-ns accuracy}
\label{sec:detTiming}
The synchronization between our signal and the bright SLR pulses allowed us to predict the expected time of arrival $t_\mathrm{ref}$ of the photons retroreflected by the satellites, which is not periodic along the orbit due to the terminal's motion. In fact, the time difference between two consecutive $t_\mathrm{ref}$ deviates by a factor $1+2\frac{v_\mathrm{r}}{c}$ where $v_\mathrm{r}$ is the satellite radial velocity and $c$ is the speed of light.

As described in Section~\ref{sec:Setup}, the SPAD was used at the ground station to detect the incoming quantum signal. The detector has a characteristic temporal response $f(t)$ given by a Gaussian distribution followed by an exponential decay function
\begin{equation}
	\begin{aligned}
    f(t) =  A \mathrm{e}^{-\frac{(t-t_0)^2}{2\sigma^2} } \Theta(t_1-t) + B \mathrm{e}^{-\frac{t-t_1}{\tau}} \Theta(t-t_1)
 \label{eq:detResp}\ ,
    \end{aligned}
\end{equation}
where $\sigma$ is the Gaussian standard deviation, $t_0$ is the Gaussian peak position, $t_1$ the crossover between Gaussian and exponential trend, $\tau$ is the exponential decay constant,  $\Theta(t)$ is the Heaviside function, $A$ is the Gaussian peak value and $B=A \mathrm{e}^{-\frac{(t_1-t_0)^2 }{2\sigma^2}}$ is the function value at the crossover point~\cite{Giudice2007}. 

The detection events  tagged by the TDC  give a measurement of the effective time of arrival $t_\mathrm{meas}$ of the photons. By calculating the time differences between the expected and measured time of arrivals $\Delta = t_\mathrm{meas} - t_\mathrm{ref}$ we can obtain the detection histograms reported in Fig.~\ref{fig:detectionHist}.

\begin{figure*}[t]
\centering\includegraphics[width=\textwidth]{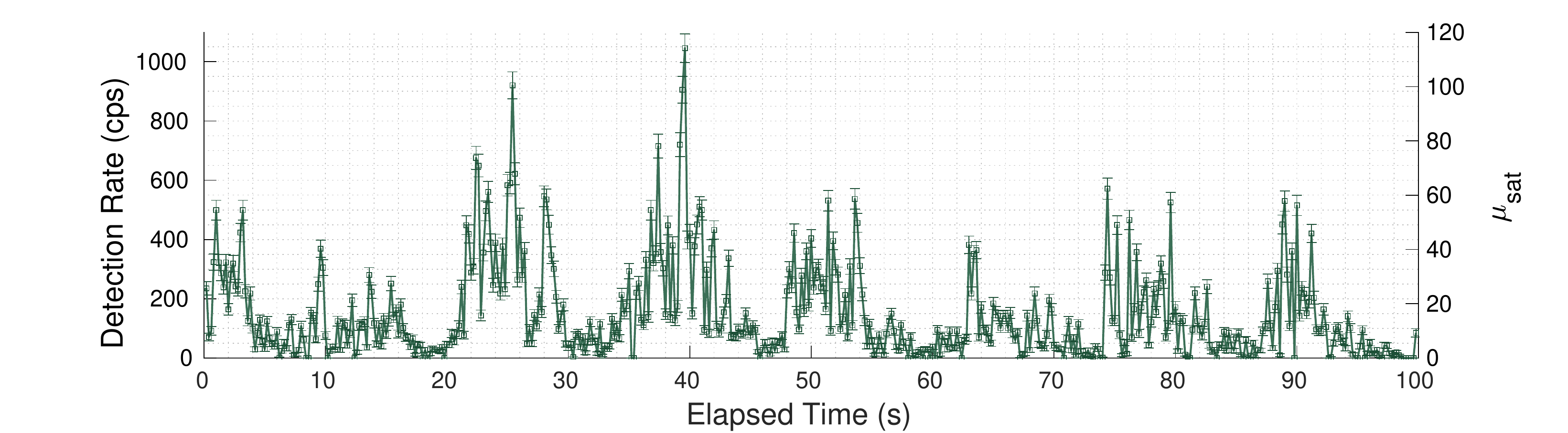}
\caption{Instantaneous detection rate and $\mu_\mathrm{sat}$ calculated every 200~ms for a 100~s sample of LAGEOS-II passage. By discarding frames with few detection events, an  average to detection rate $\bar{R}_\mathrm{det}  \approx 210$~cps with a SNR $\approx 7$ was observed. A mean number of photons at the satellite $\bar{\mu}_\mathrm{sat} \approx 16$ is observed. Since $R\approx8200$~km can be considered constant in 100 s, $\mu_\mathrm{sat}$ is related to the detection rate by a constant multiplicative factor (see Eq.~\eqref{eq:muSat}). }
\twocolumngrid
\label{fig:detectionRate}
\end{figure*}

In Fig.~\ref{fig:detectionHist}(a), the detection histogram for 5~s of LAGEOS-I passage can be observed. By fitting the calculated time differences between the expected and measured time of arrivals $\Delta$ with Eq.~\eqref{eq:detResp} the estimates $\sigma \approx 60$~ps, $\tau \approx 200$~ps and a FWHM$\approx 230$~ps are obtained. This corresponds to the convolution between the response functions of the SPAD (with typical FWHM of 65~ps) and of the time-to-digital converter with the temporal profile of the incoming quantum signal. Similar results were also obtained in the passage of LAGEOS-II. It is worth noticing that the FWHM  measured with our detection system is compatible with the impulse response of the LAGEOS satellite computed in~\cite{Degnan1993}.

As a comparison, the time differences between the expected and measured time of arrivals $\Delta$ for 5 seconds of the Beacon-C  passage can be observed in Fig.~\ref{fig:detectionHist}(b). It is clear that the temporal profile is broadened with respect to the one observed with LAGEOS-I. This can be explained by the pyramid trunk shape of the satellite which gives rise to an optical transfer function that effectively spreads the pulse duration when viewed at non-normal incidence. On the contrary, the spherical shape of LAGEOS-I avoids such pulse spreading~\cite{Degnan1993}. By fitting the data with Eq.~\eqref{eq:detResp} a FWHM $\approx 510$~ps is obtained for Beacon-C.

\subsection{Detection Rate}
\label{sec:detRate}
Now we present the evaluation of the detection rate achievable with the timing accuracy described above. We divided a 100-s sample of a passage of LAGEOS-II into frames of 200~ms  and for each frame we obtained the detection histograms as described in Section~\ref{sec:detTiming}. Then, we applied a temporal filter of 400~ps centered around the peak to discriminate between the signal and background photo detection events. In Fig.~\ref{fig:detectionRate} the detection rate obtained in the 100-s sample can be observed. This detection rate was calculated as the number of events within the temporal filter after subtracting the background counts, which  were estimated from the detections outside the temporal filter.

\begin{figure}
\centering\includegraphics[width=0.9\columnwidth]{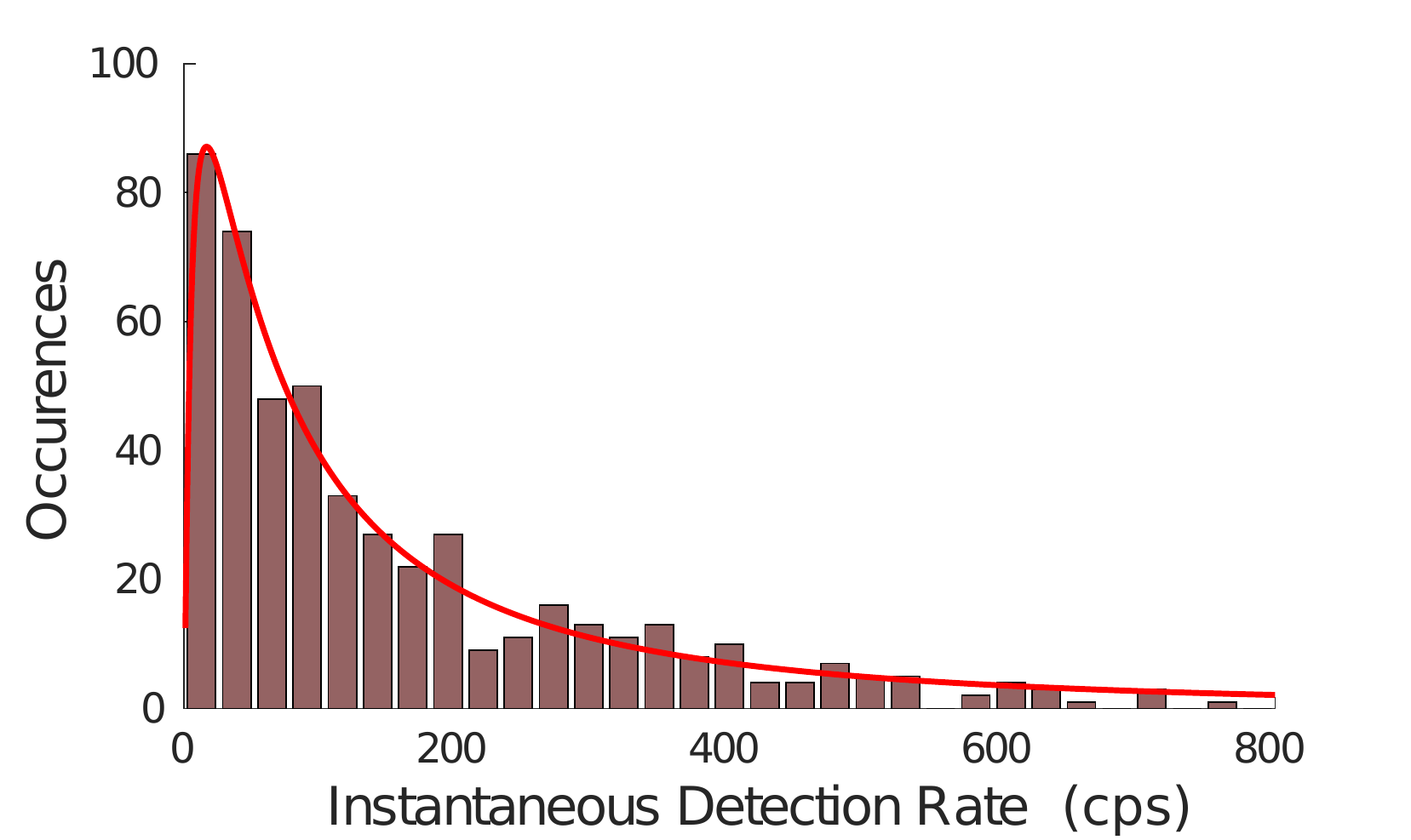}
\caption{Histogram of the occurrences of the instantaneous detection rate. The bin size of the histogram is 25~Hz.  A lognormal distribution (see Eq.~\eqref{eq:lognormal}) with a logarithmic mean  detection rate $\mu = 4.7 \pm 0.1$, logarithmic standard deviation $\sigma = 1.4 \pm 0.1$   and a scintillation index $\mathrm{SI} =  6.1 \pm 1.8$  is obtained by fitting the data.}
\label{fig:detectionRateHist}
\end{figure}

The detection rate averaged to $\bar{R}_\mathrm{det} \approx 160$~counts per second (cps) with peaks of up to 1~kcps, and an average SNR~$\approx 5$. If we discard all frames with an SNR less than 1 (about 25\% of the total), the detection rate  averaged to $\bar{R}_\mathrm{det}  \approx 210$~cps with an SNR~$\approx 7$. Such post-selection is performed to counteract the high quantum bit error rate (QBER) caused by turbulence and scintillation  in the spirit of an Adaptive Real Time Selection method for QKD~\cite{Vallone2015_ARTS}. In fact, turbulence and scintillation causes a fluctuation of the signal rate while the background noise remains constant. This translates into a fluctuation of the SNR which has negative repercussions on the QBER. 
{  It is important to note that the large fluctuations seen in Fig.~\ref{fig:detectionRate} are mainly caused by atmospheric turbulence in the ground-to-satellite link. This can be attributed to a turbulence-induced beam wandering that is similar in magnitude to the beam size during the up-link propagation. Instead, for satellite-to-ground link, the beam size is much larger than the turbulence-induced beam wandering, leading to  smaller fluctuations, as reported by S.-K.~Liao~\emph{et al.}~\cite{Liao2017}.}

The histogram of the occurrences of the instantaneous detection rate can be observed in Fig.~\ref{fig:detectionRateHist}. The histogram is obtained with a binning of 25~Hz. The data is fitted with a lognormal (LN) distribution
\begin{equation}
P_\mathrm{LN}(x;\mu, \sigma) = \frac{1}{x\sigma\sqrt{2\pi}}  e^{-\frac{\left(\ln x-\mu\right)^2}{2\sigma^2}} \ , \label{eq:lognormal}
\end{equation} 
with a logarithmic mean  detection rate $\mu = 4.7 \pm 0.1$ and logarithmic standard deviation $\sigma = 1.4 \pm 0.1$.
The fitting curve was chosen since it is expected that turbulence gives rise to a fluctuating channel with a transmissivity varying according to a lognormal distribution~\cite{Capraro2012, Vasylyev2016}.
From the fitted parameters a scintillation index $\mathrm{SI} = \frac{\Delta x^2}{\left\langle x \right\rangle^2} = e^{\sigma^2} -1 =  6.1 \pm 1.8$ can be estimated.
However, further analysis are necessary to confirm such hypothesis with general ground-to-satellite-to-ground channels.

\subsection{Determining  the mean photon number per pulse at the satellite}
\label{sec:muSat}
The mean number of photons reflected by the satellite per pulse can be determined as the average number of received photons per pulse at the ground station $\mu_\mathrm{rec}$ divided by the downlink part of the radar equation~\cite{Degnan1993}. After having evaluated $\mu_\mathrm{rec}$ from the instantaneous detection rate divided by the number of sent pulses, the mean number of photons reflected by the satellite is given by
\begin{equation}
    \mu_\mathrm{sat} = \frac{\mu_\mathrm{rec}} {T_\mathrm{diff} T_\mathrm{A}(R) \eta_\mathrm{rx} \eta_\mathrm{det}} \ ,
 \label{eq:muSat} 
\end{equation}
where $T_\mathrm{diff}$ is the diffraction transmittance, $T_\mathrm{A}(R)$ is the atmospheric transmittance at a given distance $R$~\cite{Degnan1993}, $\eta_\mathrm{rx} = 0.13$ is the transmission of the whole receiving apparatus and $\eta_\mathrm{det} = 0.5$ is the detector efficiency. The diffraction transmittance can be estimated in the top-hat approximation for the Far Field Diffraction Pattern with solid angle $\Omega$~\cite{Dequal2016} as $T_\mathrm{diff} = \frac{A_\mathrm{tel}}{\Omega R^2}$,   with $\Omega$   given by~\cite{Degnan1993}
\begin{equation}
   \Omega = \frac{4 \pi A_\mathrm{CCR} \rho N_\mathrm{eff} }{\Sigma} \ ,
 \label{eq:solidAngle} 
\end{equation}
where $\Sigma = 15\times10^6\mathrm{m}^2$~\cite{Arnold2003} is the array cross-section, $A_\mathrm{CCR} = 11.4\times10^{-4}\mathrm{m}^2$~\cite{Arnold1978} is the CCR reflective area, $\rho = 0.89$~\cite{Minott1993} is the CCR reflectance at normal incidence and $N_\mathrm{eff} = 9.88$~\cite{Arnold1978} is the effective
number of CCRs averaged over all orientations. It is useful to note that the Far Field Diffraction Pattern with solid angle $\Omega$ for the LAGEOS satellites corresponds to a source with ${\sim}100$~$\mu$rad of angular aperture and ${\sim}55$~dB of diffraction losses.

From the data obtained in the 100-s sample of LAGEOS-II, a mean number of photons at the satellite $\bar{\mu}_\mathrm{sat} \approx 16$ is calculated. The instantaneous $\mu_\mathrm{sat}$ can be observed in Fig.~\ref{fig:detectionRate}. Since in 100~s the distance $R\approx8200$ 
~km can be considered constant, $\mu_\mathrm{sat}$ is related to the detection rate by a constant multiplicative factor.

\section{Discussion}
Here we reported that, by exploiting SLR and using a commercially available SPAD and TDC, a temporal accuracy of $230$~ps in the detection of an optical signal of few photons retro-reflected by LAGEOS   was obtained. Furthermore, we observed a mean detection rate of $\bar{R}_\mathrm{det}  \approx 210$~cps with a SNR~$\approx 7$. If the retroreflectors of LAGEOS satellites were replaced by an active source with $10$~$\mu$rad of angular aperture~\cite{Liao2017}, shrinking the beam spot on ground, and   by using the point-ahead technique to compensate for velocity aberration~\cite{Han2018}, it would be possible to reduce the diffraction losses by 20~dB with respect to the mean channel losses observed in this study. The receiving ground station could also be improved by removing the two BS used to separate the outgoing and incoming beams of our scheme and by performing SQC and SLR with different wavelengths. Such modifications would
avoid signal losses due to beam splitters which accounts to 6~dB. 
The transmitter in MEO orbit with improved divergence, 100~MHz repetition rate source and $\mu_\mathrm{sat} \approx 1$, together with the improved ground station with $\eta_\mathrm{rx}\approx 1$, could allow for a detection rate of approximately 10~kcps with a high SNR around 350.

The sub-ns temporal accuracy here reported demonstrates that QKD implementations with GHz repetition rates are compatible with SQC. Furthermore, this timing accuracy can allow for time-bin encoding with shorter temporal imbalances than the one previously reported~\cite{Vallone2016, Vedovato2017}. Reducing the temporal imbalances would allow the use of  more stable interferometers, for example by exploiting the birefringence of calcite~\cite{Calderaro2018_WM}, hence resulting in increased visibility, and a lower kinematic phase modulation due to the satellite motion, which can be estimated via SLR and actively compensated. Lastly, it would pave the way for the implementation of multidimensional time-bin encoding \cite{Islam2017} for SQC.

%\section*{Funding}
%{\red Please identify all appropriate funding sources by name and contract number. Funding information should be listed in a separate block preceding any acknowledgments. List only the funding agencies and any associated grants or project numbers, as shown in the example below:

%National Science Foundation (NSF) (1253236, 0868895, 1222301); Program 973\\ (2014AA014402); Natural National Science Foundation (NSFC) (123456).

%OSA participates in \href{https://www.crossref.org/fundingdata/}{Crossref's Funding Data}, a service that provides a standard way to report funding sources for published scholarly research. To ensure consistency, please enter any funding agencies and contract numbers from the Funding section in Prism during submission or revisions.}

\section*{Acknowledgments}
We would like to thank Francesco Schiavone, Giuseppe Nicoletti, and the MRLO technical operators for the collaboration and support.
We acknowledge the International Laser Ranging Service (ILRS) for the satellite data. 
Our research was partially funded by the Moonlight-2 project of INFN.

%%%%%%%%%%%%%%%%%%%%%%%%%%  references  %%%%%%%%%%%%%%%%%%%%%%%%%%
\bibliographystyle{apsrev4-1}
\bibliography{main}

\end{document}